\documentclass[apj]{emulateapj}
\input epsf
\begin{document}
\def\beq{\begin{equation}}
\def\eeq{\end{equation}}
\def\bey{\begin{eqnarray}}
\def\eey{\end{eqnarray}}
\def\pc{\, {\rm pc} }
\def\kpc{\, {\rm kpc} }
\def\msun{M_\odot}
\def\sun{\odot}
\def\lsim{\mathrel{\raise.3ex\hbox{$<$\kern-.75em\lower1ex\hbox{$\sim$}}}}
\def\gsim{\mathrel{\raise.3ex\hbox{$  $\kern-.75em\lower1ex\hbox{$\sim$}}}}
\def\Msun{M_\odot}
\def\Lsun{L_\odot}
\def\lsun{L_\odot}
\def\kms{\, {\rm km \, s}^{-1} }
\def\eV{\, {\rm eV} }
\def\keV{\, {\rm keV} }
\def\dis{{\rm dis}}
\def\grad{{\bf \nabla}}
\author{HongSheng Zhao\altaffilmark{1} \& Baojiu Li\altaffilmark{2}}
\altaffiltext{1}{Scottish University Physics Alliance, University of St Andrews, KY16 9SS, UK, hz4@st-andrews.ac.uk}
\altaffiltext{2}{Department of Applied Math and Theoretical Physics, Cambridge University, CB3 0WA, UK, b.li@damtp.cam.ac.uk}
\title{Dark Fluid: Towards a unification of empirical theories of galaxy rotation, Inflation and Dark Energy}
\begin{abstract}
Empirical theories of Dark Matter like MOND gravity and of Dark Energy like f(R) 
gravity were motivated by astronomical data. But could these theories be 
branches rooted from a more general hence natural framework? Here we propose the
natural Lagrangian of such a framework based on simple dimensional analysis and
co-variant symmetry requirements, and explore various outcomes in a top-down
fashion. Our framework preserves the co-variant formulation of GR, but allows the 
expanding physical metric be bent by a single new species of Dark Fluid flowing in 
space-time. Its non-uniform stress tensor and current vector are simply functions of 
a vector field of variable norm, resembling the 4-vector
electromagnetic potential description for the photon fluid, but is dark (e.g.,
by very early decoupling from the baryon-radiation fluid). The Dark Fluid
framework naturally branches into a continuous spectrum of theories with Dark
Energy and Dark Matter effects, including the $f(R)$ gravity, TeVeS-like
theories, Einstein-Aether and $\nu\Lambda$ theories as limiting cases. When the
vector field degenerates into a pure Higgs-like scalar field, we obtain the
physics for inflaton and quintessence. In this broad setting we emphasize the 
non-constant dynamical field behind the cosmological constant effect, and 
highlight plausible corrections beyond the classical MOND predictions. 
Choices of parameters can be made to pass BBN, PPN, and causality constraints. 
The Dark Fluid is inspired to unify/simplify the astronomically successful 
ingredients of previous constructions: the desired effects of inflaton plus 
quintessence plus Cold DM particle fields or MOND-like scalar field(s) are shown 
largely achievable by one vector field only.
\end{abstract}

\maketitle

\section{Introduction}

Gravity, the earliest and the weakest of the known forces, has
never been very settled.  The beauty of co-variant symmetry motivated
Einstein to supersede the Newton's paradigm with General Relativity, 
but (inadequate) empirical evidences motivated Einstein
to introduce first and then withdraw the cosmological constant,
a concept defying quantum physics understanding even in modern day.
While making generally tiny or a factor of two corrections, 
the equivalence principles insist on certain
symmetries in space-time, \emph{e.g.}, co-variant symmetry and no frame 
to measure locally any absolute direction of gravitational acceleration for a free-falling observer.  
However, symmetry can be spontaneously broken if there are dynamical
interactions or couplings of fields; a well-known mechanism in
several branches of physics, especially the Higgs-mechanism in
particle physics to give a mass to a particle. Many attempts have
been made to break the strong equivalence principle by adding new
fields (degrees of freedom) in the gravity sector, which
essentially means the gravitational "constant" $G$ may be is a new
dynamical degree of freedom governed by other fields coupled to
the metric.  The best known is the Brans-Dicke theory (1961). The
lesser known is that a vector field of non-zero absolute value 
in vacuum can also be coupled to gravity, to give absolute directions (Will 1993). 
It has long been suggested that Lorentz symmetry can be broken locally in the quantum
gravity and string theory context (Kostelecky \& Samuel 1989) to yield a vector
field of a non-zero expectation value (\emph{e.g.},
pointing towards the direction of time) in the vacuum. The most
successful attempt so far is the Einstein-Aether theory of
Jacobson et al. (2001).
A common theme of these theories is that they are \emph{not}
invented for certain observational anomaly. Rather in the same
vein as how symmetry motivated General Relativity, these theories
meet the astronomical data only a posteri, e.g.,  Li, Mota, Barrow
(2007) showed a vector field in the gravity sector could NOT be excluded by 
the accurate Cosmic Microwave Background (CMB) data.

Nevertheless, the above order is not the only way to discover  
theories. The puzzling black body radiation spectrum and Balmer's
curious empirical formula for hydrogen lines are among the odd
pieces of classical physics, which lead to full formulation of quantum mechanics. 
This bottom-up approach is often gradual, the arrival of the final theory
taking several generations of formulations (\emph{e.g.}, from 
Planck's model for blackbody radiation and Bohr's model for hydrogen atom to
Heisenberg's matrices-based formulation in general) with different levels of mathematical
rigour and sophistication.

Historically, Milgrom's MOdified Newtonian Dynamics
(MOND) was invented without any packaging by co-variant theories of gravity,
just as the Dark Matter empirical concept was invented by Zwicky without 
packaging first with SuperSymmetry-like particle field theories.
MOND is a formula motivated to reveal the curious uniform rules (or facts) 
underlying rotation curves of many spiral galaxies, as Balmer's
formula and its generalisations suggesting strongly a fundamental
rule for all atomic lines. Since the rule is empirical and bare
(without co-variance), it waits to be enshrouded by a theory
preserving basic symmetries to predict any logical corrections to
situations where the empirical rule must fail slightly,
\emph{e.g.}, by factor of two in some gravitationally lenses made
by elliptical galaxies and clusters of galaxies.  The
Tensor-Vector-Scalar (TeVeS) framework of Bekenstein (2004),
unifying earlier constructions by Sanders and others, makes the
first step to the integration of MOND formula with fundamental
physics.  A time-like vector field is shown to be 
the necessary ingredient of a MOND gravity.
Yet the original aim of TeVeS was limited, \emph{e.g.}, not
addressing the cosmological constant problem, or the inflation.
Orthogonally many literatures considered theories of modified
gravity such as the $f(R)$ gravity (Chiba 2003) and scalar inflation theory as
\emph{ad hoc} fixes of the cosmological constant problem and the
horizon problem respectively, without aiming to address outstanding
questions on galaxy rotation curves.  
Recently Zhao (2007), built on the work of Zlosnik et al. (2007), showed
that these outstanding problems of DM and DE can find at least one common
solution simultaneously in the framework of a massive vector
field.  In these theories, there is 
"One Field which rules them all and in the darkness bind them."

The most famous examples of a vector field is the massless spin-1
photon and the massive Z-boson in the electro-weak theory. The
standard way to give masses to particles in particle physics is
the Higgs mechanism where a scalar field, coupled to the vector
field, acquires a non-zero value in vacuum. Turning to the gravity
sector, however, it is unclear how the fundamental prediction of
spontaneous symmetry breaking from quantum gravity might be
related to the mundane effects of galactic dark matter and
cosmological constant.  We lack a framework.

We shall show that MOND, $f(R)$ gravity, Einstein-\AE ther theory
and inflation can be integrated into a common framework with a
unit vector and a dynamical scalar field. MOND would
become a specific choice of the potential of the scalar field.
Having such a framework allows one to explore the consequences of
modified gravity systematically. It can be meaningless to even
differentiate dark energy and modified gravity.
Modified gravity contains extra fields, which can be
treated as dark energy field.

The goal of this paper is to show the existence of a very
general Lagrangian for which the MOND formulae are the natural consequences
in spiral galaxies in equilibrium, rather than the golden rule for
(non)-equilibrium systems of all scales.  We demonstrate this with the modified Poisson
equation and for the equation for the Hubble expansion.

The outline is as follows. We propose our general Lagrangian in \S2, and
illustrate how it reduces to various special cases, TeVeS, BSTV,
Einstein-Aether, $f(K)$, $f(R)$, inflation.  We choose a subset of 
models with MOND and Dark Energy effects in \S3.  We examplify the
properties of our dark fluid in the case of
Hubble expansion and inflation (\S4), and for static galaxies (\S5). We discuss
corrections to MOND in \S6, and summarize the properties of the Dark Fluid in \S7.
Appendix gives an estimate of the damping frequency of the Dark Fluid.

\section{The proposed Lagrangian for the Dark Fluid}

Denote a vector field by $Z^a$, which has generally a variable or
dynamic norm
\beq
\varphi^2 \equiv g_{ab} Z^aZ^b,
\eeq
which is essentially an auxiliary scalar field characterizing the
norm of the vector field $Z^a$, hence is not an independent
dynamical freedom.  A generic coupling of the vector
field $Z^a$ with the space-time is through the contractions among
the $Z^aZ^b$ tensor, the $g_{ab}$ metric tensor, the Ricci tensor
$R_{ab}$.  Hence the most generic theory of the vector
should contain the terms
\begin{eqnarray}\label{glagv}
L &=& \left[g^{ab} + g^{ab} f_1\varphi^2 - f_2 Z^{a} Z^{b}\right]R_{ab} + f_{34}, \\
f_{34} &\equiv &  f_3 \varphi^2 + f_4 \nabla^{a}\varphi\nabla_{a}\varphi + \cdots
\end{eqnarray}
where $f_i$ could be constants or functions of $\varphi$. 
The term $f_2 Z^a Z^b R_{ab} = 
f_2 Z^{a}\left(-\nabla_{a}\nabla_{c}Z^{c} + \nabla_{c}\nabla_{a}Z^{c}\right)$
could always be recast through a full divergence to
$\rightarrow K^{ab}_{cd} \nabla_a Z^b \nabla_c Z^d$ 
where $K^{ab}_{cd}$ are constant tensors or tensor functions.  
The equivalent expression for the Lagrangian is
\begin{eqnarray}\label{fullderiv}
L & = & K^{ab}_{cd} \nabla_a Z^b \nabla_c Z^d + \Theta_0 R +
\Theta_1 \varphi^2 + \Theta_2 \nabla^{a}\varphi\nabla_{a}\varphi + \cdots
\end{eqnarray}
where $R=g^{ab}R_{ab}$ is the Ricci scalar, and
$K^{ab}_{cd}$ are constant tensors for simplicity but can
also be lengthy functions of $\varphi$ in general; $\Theta_i$ are
constants in the simplest case but functions of $\varphi$ in
general.

The above Lagrangian is generic enough, and many dark energy
models can be derived from it. For example, the terms
$(1+\varphi^2) R + \omega \varphi^{2/(1+n)}$ can lead to an $R +
\mathrm{const}/R^{n}$ (Li \& Barrow 2007) gravity (as could be
checked by solving $\varphi$ from the equation of motion of
$\varphi$ and then substituting back into the original
Lagrangian); and the terms $ \phi R - 4 
(1-\phi^{-1}) \omega \nabla^{a}\varphi\nabla_{a}\varphi$ leads to
the Brans-Dicke theory of gravity, where the auxilliary field
$\phi=1+\varphi^{2}$. The usual inflation-like or quintessence
theories can be recovered from the terms $ R + V(\varphi) +
\nabla^{a}\varphi\nabla_{a}\varphi$.

The essential dynamics of a cosmological vector field
is described by the first term
$K^{ab}_{cd} \nabla_a Z^b \nabla_c Z^d$
(Ferreira et al. 2007, Halle, Zhao \& Li 2008). If insisting that the field
has a unit norm guaranteed by a Lagrange multiplier, one would
recover Einstein-\AE ther theory (Jacobson et al. 2001)
and its generalizations (Zlosnik \emph{et al.} 2007, Li \emph{et al.} 2007).

\section{MOND-inspired subset of dark fluid theories}

To see the relation to MOND, consider the subset where
$\Theta_0=1$, $\Theta_1 \sim \Lambda_0$, and $\Theta_2 \sim N^{-2}$.
Let's decompose the four dynamical freedoms in the vector field
into a unit norm part of 3 degrees of dynamical freedom, 
\beq
\AE^a \equiv {Z^a \over \varphi}, \quad \lambda \equiv \varphi^2 \equiv
g_{ab}Z^aZ^b ,  
\eeq 
and rewrite the scalar field $\varphi$ in term of the new scalar field $\lambda$.
The Lagrangian is then casted to a form containing {\it at least}
the following
\begin{eqnarray}
\mathcal{L} &=& L_m + R + L_\varphi + L_{\AE} \\
L_\varphi &=& - {1-c_\varphi^2 \over N^2} \nabla_\parallel \varphi \nabla_\parallel\varphi
- {c_\varphi^2 \over N^2} \left(\nabla_a \varphi\right)\left(\nabla^a \varphi\right) + 2\Lambda_0 F(\lambda)
+ \cdots, \\
L_{\AE} &=& c_4 \nabla_\parallel\AE_{c}
\nabla_\parallel\AE^{c} + c_2 ~ (\nabla_{a}\AE^{a})^2 
+ (\AE^{a}\AE_{a}-1)L^* + \cdots,
\end{eqnarray}
where $N$, $F$, $c_\varphi^2$, $c_2$ and $c_4$ are various coupling constants or functions 
of $\lambda \equiv \varphi^2$, and 
$\nabla_\parallel\equiv\AE^{a}\nabla_{a}$, 
$\nabla^a=g^{ac}\nabla_c$, $\AE_a=g_{ac}\AE^c$,
$R$ is the Ricci scalar, and $L^*$ is the Lagrange multiplier (a kind of potential).
The Lagrangians $L_m$, $L_\varphi$, $L_{\AE}$ are for the
matter, the scalar field $\varphi$ and the unit vector field
$\AE^a $ respectively, where we omitted two possible terms 
$c_1 K_1+c_3 K_3 = c_1 (\nabla_{a}\AE_{b}) (\nabla^{a}\AE^{b}) 
+ c_3 (\nabla_{a}\AE_{b}) (\nabla^{b}\AE^{a})$ in $L_{\AE}$.
$\Lambda_0$ is the only dimensional scale in the dark fluid, 
it is a scale of energy density.  

In this Lagrangian we have three dynamical fields: the scalar
field $\lambda$, the \AE ther field $\AE^{a}$ and the metric field
$g^{ab}$ plus a non-dynamical $L^*$.  
Now varying the action $S= - \int \sqrt{-g}d^4 x {L \over 16 \pi G}$ 
with respect to them will lead to 
the scalar field equation of motion (EOM), \AE ther field EOM and
the modified Energy-momentum tensor plus the unit vector constraints for $\AE$ field respectively. The general
results are more tedious and are presented elsewhere (Halle, Zhao,
Li 2008). Here we illustrate the physics by considering only the
main terms for a specific choice of functions.

\subsection{Choices of coupling constants}

The dimensionless function $F(\lambda)$ has the meaning of the potential of 
the scalar field $\lambda \equiv \varphi^2$, 
and $c_\varphi^2$, $c_2$ and $c_4$ are of order unity, and 
are also generally functions of $\lambda$ and 
will be shown to be related the sound speeds of the Dark Fluid.  

\begin{itemize}
\item We set the scalar field sound speed 
\bey
c_\varphi &\sim & 1,  
\eey 
and we shall treat $N$ and $c_\varphi$ as constants.

\item We choose the coefficients
\bey
{c_4(\lambda) \over 2} &=& \lambda \equiv \varphi^2\\ 
{c_2(\lambda) \over 2} & \equiv & {b(\lambda) -1 \over 3} = -{1 \over 3 \lambda},
\eey
i.e. $b=1-\varphi^{-2}$.  
For simplicity we set two other terms in $L_{\AE}$ of Jacobson's unit vector
field to zero, i.e., $c_1 K_1 =0$ and $c_3 K_3=0$.  This might not be necessary, but 
simplifies the analysis of PPN parameters 
in the solar system and the sound speed of the vector field.
Our choice of Lagrangian with $c_1=c_3=0$ kills 
spin-1 mode waves of the vector field, and gurantees that the normal gravitational wave
in the tensor mode will propagate with the normal speed (of light).
This choice is perhaps not necessary, but is intended to avoid contraversy on the causuality issue.  
Even the spin-0 mode sound speed $c_{\AE}$ is plausible: 
a rigourous analysis (Foster \& Jacobson 2006, valid for any constant $\lambda$ and $b$) 
predicts  
\beq
c_{\AE}^2 = {c_2 \over c_4}{(2-c_1) \over (2 + 3 c_2)} 
= {(1-b) (\lambda-1)\over 3 b \lambda} ={1 \over 3\varphi^2} >0.
\eeq
Interestingly in the solar system, where $\lambda \equiv \varphi^2 \rightarrow 0$, 
the spin-0 mode of the vector field propagates almost instantaneously
with $c_{\AE}$ being $(3\lambda)^{-1/2}$ times bigger than the speed of light,  
avoiding the Cherenkov radiations constraint in the solar system.
All PPN parameters are expect to be equal to that of GR in the solar system as well;    
although the PPN parameters $\alpha_1 = -8\lambda $, $\alpha_2 = (3\lambda -1)\lambda$ 
are non-zero, as shown later the scalar field 
$\lambda$ is expected to settle to a very small 
equilibrium value $\sim (10^{-10})^4$ on earth for our choice of the 
penalizing scalar field potential $F(\lambda)$.

\item Our choice for the dark fluid potential (shown in Fig.1) is 
\bey 
\Lambda_0 F(\lambda)|_{\lambda=\varphi^2} &=&  {8 (\varphi-1)^3 \over 3}\Lambda_0.
\eey
The dark fluid's energy scale is defined by $\Lambda_0$.  An important property is that
in the limit $\lambda=\varphi^2 \rightarrow 1$, we have 
\bey
F' \equiv {d \over d\lambda}F 
& \sim & (1-\lambda)^2, ~{\rm if}~ \lambda=\varphi^2 \rightarrow 1,\\ 
& \propto & \lambda^{-1/2}, ~{\rm if}~ \lambda=\varphi^2 \rightarrow 0;
\eey
note a prime always means ${d \over d\lambda}$).  
We shall show that this property 
describes {\it a non-uniform (dark energy) fluid which gives 
the MOND-like (dark matter) effects in galaxies but not in the solar system}.
\footnote{A more general choice 
of dark fluid potential of this property is  $F \propto \int (\varphi^n-1)^2 d\varphi$
for $n=1$ (as above) and $n=2,3,4, ...$.  These potentials are always simple polynomial functions of $\varphi$.}   
\end{itemize}

\section{Background cosmology}

Consider background cosmology in the FRW flat metric,
\begin{eqnarray}
ds^{2} &=& dt^{2} - a^{2}(t) (dx^2 + dy^2 + dz^2).
\end{eqnarray}
Firstly, the scalar field follows an equation of motion exactly as
a quintessence,
\begin{eqnarray}
\ddot{\varphi} + 3H\dot{\varphi} =-
 \left(\Lambda_0 F' + 3b'H^{2}\right) (2N^2\varphi),
\end{eqnarray}
so $\varphi$, the norm of the vector field $Z^a$, tracks the
Hubble rate $H=\dot{a}/a$. The vector field equation of motion
gives the Lagrange multiplier (or the mass of the vector field)
$L^* = \partial_t (\alpha H) + {1-c_\varphi^2 \over N^2}\dot{\varphi}^{2}$, where
$\alpha=2b-2$. This mass is varying with time, hence the vector
field describes effectively an unstable slowly decaying particle.

The modified 00-term of the Einstein equation becomes
\begin{eqnarray}
3 H^2 &=& 8\pi G \left[ \bar{\rho} + \bar{\rho}_{DM} + \bar{\rho}_{DE} \right], \\
\bar{\rho}_{DM} &\equiv &(b^{-1}-1) \bar{\rho} \\
\bar{\rho}_{DE} &\equiv& {1 \over 8\pi G b}\left[{1\over 2N^2}\dot{\varphi}^{2} + \Lambda_0 F \right],
\end{eqnarray}
where $\bar{\rho}$ is the (background) energy density of
baryon-radiation fluid.  For our choice of $b$, 
\beq
b^{-1} = {\varphi^2 \over \varphi^2-1},
\eeq
so we get a 
dark-matter-like effect of the vector field for $1 < \lambda=\varphi^2 \le 2$
by amplifying the gravitational constant $G$ by a factor $\infty > b^{-1} \ge 2$. 
There is no dark matter-like effect at very high redshift, e.g., radiation era or BBN,  where $\lambda \rightarrow \infty$, hence $b \rightarrow 1$.  So the BBN constraint
is automatically satisfied because the Hubble expansion rate 
at BBN is equal to that of a radiation only universe. 

Equivalently the Einstein equation can be written as
\begin{eqnarray}
 -\left( 2\frac{\ddot{a}}{a} + H^{2}\right) &=& 8\pi G
 \left(\bar{p} + \bar{p}_{DM}
+ \bar{p}_{DE}\right)\\
\bar{p}_{DM} &=& (b^{-1}-1) \bar{p}  \\
\bar{p}_{DE} & \equiv & {1 \over 8\pi Gb}\left[{1 \over
2N^2}\dot{\varphi}^{2} - \Lambda_0 F + 4b'H\varphi\dot{\varphi}
\right],
\end{eqnarray}
where the pressure of the baryon-radiation fluid $\bar{p}=0$ in
the matter-dominated era, hence $\bar{p}_{DM}=0$. Apply the slow-roll approximation in the
late universe we find the effective pressure of the vector field
\beq
-\bar{p}_{DE} \sim \rho_{DE} \sim {\Lambda_0  \over 8\pi G b} F
\sim {\Lambda_0 \over 3\pi G} {(\varphi-1)^2 \varphi^2\over (\varphi+1)}.
\eeq 
This behaves like a dark 
energy with \beq w = \bar{p}_{DE}/\bar{\rho}_{DE}  \sim -1 \eeq
and with a characteristic scale $\Lambda_0$, which must be
set of order $(8 \times 10^{-10}m/s^2)^2$ to match the observed
cosmological constant. Note that in writing above equations we have
implicitly assumed that the effective dark matter and effective dark energy components
couple to each other. This can been seen by checking that neither
$\bar{\rho}_{DM}$ nor $\bar{\rho}_{DE}$ satisfies the conservation
law $\dot{\bar{\rho}} + 3H(\bar{\rho}+\bar{p})=0$, but their sum
does. The coupling strength is determined by $b'$: if $b$ is a 
constant (equals to unity for our model in the early universe), 
then the two components decouple. 

Models with a very large $N$ could even drive 
inflation in the early universe when the energy density is
dominated by the norm of the vector field (\emph{i.e.}, the scalar
field $\varphi$). It is easy to verify the solution 
\beq 
\Lambda_0 {8 \varphi^3 \over 3} \sim 3 H^2 b, \quad 
{d\varphi^2 \over d\ln a} \sim -6 N^2 b^{-1}, \eeq applies 
in the slow-roll phase, where $b \sim 1$ for very large $\varphi$.  
This phase of slow rolling can inflate the universe 
by a factor $\sim \exp\left({\varphi_i^2 - \varphi_f^2 \over 6 N^2}\right)$ 
(Kanno \& Soda 2006).  
For the universe to inflate by a factor $\exp(60)$, \emph{e.g.}, 
$\varphi$ rolles from an initial value 
$\varphi_i \sim 20N$ to a final value $\varphi_f \sim 10N$.  
The end of inflation or the start of the radiation era is at a time
$H^{-1} \sim \left({\Lambda_0 F \over 3}\right)^{-1/2} \sim 10^{17} F^{-1/2}{\rm sec}
\sim 10^5 \left({N \over 10^{8} }\right)^{-3/2} {\rm sec}$; the time scale 
can be even shorter for other forms of Dark Fluid potential 
\footnote{e.g., 
if $F \propto \int (\varphi^4-1)^2 d\varphi \propto \varphi^9$ then $H^{-1} \sim 
10^{-19} \left({N \over 10^{8} }\right)^{-9/2} {\rm sec}$.}  
The inflation might 
end when the vector field decays partly into known particles via 
some small coupling  perhaps of the type $g_{ab}Z^a J^b$ between 
the vector field and the current field $J^b$ of some 
known fields, e.g., coupling with sterile neutrinos, which would then mix to  
neutrinos of all flavors and couple to photons, leptons and hardrons etc.  

\section{Static galaxy limit}

To work out perturbations in static galaxies, remember that in the
Newtonian gauge we have only two scalar mode perturbation
potentials, $\Phi$ and $\Psi$, which appear in the perturbed
metric:
\begin{eqnarray}
ds^{2} &=& (1+2\Phi)dt^{2}
- a^{2}_{0}(1+2\Psi) (dx^2 + dy^2 + dz^2) ;
\end{eqnarray}
we will let $a_{0}=1$.  We assume NO Hubble expansion.

The vector field equation of motion in static systems fixes
$\AE^{a}=(1-\Phi, 0, 0, 0)$, so the vector field tracks the
metric exactly without any freedom in static galaxies.

The 00-component of the Einstein equation becomes a Poisson
equation
\begin{eqnarray}
\sum_{i=x,y,z} -(2\Psi)_{,ii} & =& 8\pi G (\rho + \rho_{DM} + b\rho_{DE} ) ,\\
\rho_{DM} &\equiv& {\sum_{i=x,y,z} \over 8\pi G}
\left[2\lambda \Phi_{,i}\right]_{,i}
\end{eqnarray}
where we use notation $F_{,i} \equiv \partial_i F$, and the dummy
index implies co-variant or contra-variant derivatives with
respect to $x, y, z$. We use the approximation that the DE part
$8\pi G b \rho_{DE} = {\dot{\varphi}^2 \over 2}+\Lambda_0 F$ is a
negligible source compared to $8 \pi G \rho$ from the baryons, and
that 
\beq -\Psi = \Phi \eeq 
from the spatial cross term of the
Einstein equation. The above result is essentially a Poisson
equation where the vector field creates an effective dark
matter-like source term $\rho_{DM}$. Rearrange the terms, the same
equation becomes the MOND Poisson equation 
\beq
 \nabla \cdot \left[(1-\lambda) \nabla \Phi\right] = 4\pi G \rho,
\quad \lambda \equiv \varphi^2. 
\eeq

To see that $1-\lambda$ can be identified with the MOND $\mu_M$
function, first we define a value of the scalar field $\varphi_M$
such that \beq
 F'|_{\lambda=\varphi_M^2} \equiv  {|\nabla \Phi|^2 \over \Lambda_0 }.
\eeq We find that the scalar field equation of motion is given as
\begin{eqnarray}
-c_\varphi^2 \nabla^2 \varphi &=& -\left[ \Lambda_0 F' - |\nabla \Phi|^2
\right] (2N^2\varphi),
\end{eqnarray}
where we neglect all time-dependent terms.  This equation is similiar to 
the equation of Yukawa potential 
with a screening length of $L$, $(\nabla^2 - L^{-2}) \varphi=0$. 
In the simplest case, we adopt 
$c_\varphi^2 \rightarrow 0$ to kill the Laplacian term 
$\nabla^2 \equiv \sum_{i=x,y,z} \partial_i \partial_i $.  
In the static limit,
we find the equation for the scalar field becomes 
\beq 
{ 4\Lambda_0 } ( \lambda^{-1/4} - \lambda^{1/4} )^2 =  |\nabla \Phi|^2
\eeq
for our choice of $F(\lambda)$.  The equation can then be solved as
\beq
\lambda =\varphi^2 \rightarrow \varphi_M^2 = 
\left( \sqrt{1+ {x^2 \over 16}}+{x\over 4} \right)^{-4}|_{x = {|\nabla \Phi| \over \sqrt{\Lambda_0}}}. \eeq

To see we recover the properties of MOND function $\mu_M$ (see 
 $1-\varphi_M^2$ vs $x$ shown in Fig.1),
we rewrite the solution of the scalar field as
\begin{eqnarray}
1-\varphi_M^2&\equiv \mu_M = & \left\{%
\begin{array}{ll}
    x, & \hbox{where $x \equiv {|\nabla\Phi| \over \sqrt{\Lambda_0}} \ll 1$} \\
    1-\left({x \over 2}\right)^{-4}, & \hbox{where $|\nabla \Phi| \gg \sqrt{\Lambda_0}$} 
\end{array}
\right.
\end{eqnarray}
This is exactly the physics of MOND if 
\beq
\sqrt{\Lambda_0} \rightarrow a_0
\eeq 
is identified with the MOND acceleration scale $a_0$.  
In the solar system or strong gravity regime, the modification factor
$1-(x/4)^{-4} \sim 1$ to the 
Newtonian Poisson equation is small and reduces sharply.  In weak gravity,
applying spherical approximation around a dwarf galaxy of mass $m_{b}$, we
have $|\nabla\Phi|^2/a_0 = Gm_b r^{-2}$, and the rotation curve
$V_{cir}^2 (r) = r \nabla \Phi$. The big success of MOND in dwarf
spiral galaxies is to explain their Tully-Fisher relation $
V_{cir}^4 (r)/(G m_b) = a_0 \sim
10^{-10}m/s^2$ if $\Lambda_0 \sim (1 \times 10^{-10}m/s^2)^2$, which is  
of the order of magnitude of the observed amplitude of
"the cosmological constant" effect.  In the intermediate regime, 
our $\mu_M$ resembles the "standard" $\mu={x \over \sqrt{1+x^2}}$ function of MOND,
so it will fit rotation curves of galaxies very well.  

\section{Temporal and Spatial Corrections to MOND: Oscillations and Diffusions}

When considering merging systems like galaxy clusters,
time-dependent terms $\ddot{\lambda} \sim \dot{\lambda}^2 \sim
O(\omega^2)$ are important, where $\omega = O(|{\mathbf k}|
\sigma)$ is the inverse of the timescale to cross a system of size
$|{\mathbf k}|^{-1}$ by stars of velocity dispersion $\sigma$ in
unit of the speed of light. There can also be diffusion on small
scale due to a pressure-like term $\nabla^2\lambda = -|{\mathbf
k}|^2\lambda$.

The scalar field equation of motion becomes
\begin{eqnarray}
\left[\partial_t^2 \!\! - c_\varphi^2 \nabla^2 \!\! + (1-c_\varphi^2)\eta
\partial_t\right] \varphi &=& -2N^2\varphi\Lambda_0 \left[ 
F'(\varphi^2) - F'(\varphi_M^2)  
\right] \\
&\sim & - \left( \varphi -\varphi_M \right) \nu^2,
~ \nu^2 \equiv 4N^2\Lambda_0 F'' \varphi_M^2,
\end{eqnarray}
where $\eta^{-1}$ is a damping time scale due to coupling of $\varphi$
with the $\AE$ field (cf. Appendix), and  
the diffusion term $-c_\varphi^2 \nabla^2= c_\varphi^2 |{\mathbf k}|^2$ can be 
neglected if $c_\varphi^2 =0$.  The scalar field $\varphi$ then follows
the equation of a damped harmonic oscillator with a damping rate $(1-c_\varphi^2) \eta$
and a slightly non-linear restoring force $\sim -\nu^2\varphi$ and
an external force $\sim \nu^2\varphi_M \sim |\nabla\Phi|^2 (2N^2\varphi_M)$.
Assuming that the correction due to Hubble expansion 
\footnote{Considering the expansion of the universe would introduce 
a correction term $J b'$ in the force, where $J=(3H^2 + 2H\eta)$.}
is negligible for a very small $b'$,
the scalar field
$\varphi$ eventually approaches the MOND-like static solution $\varphi_{M}$,
thanks to the damping term with a timescale  $\eta^{-1}$, which kills 
any history dependence.  Rapid oscillations will likely keep the fluid's time-averaged
property close to MOND-like solution as well.

We estimate the oscillation time scale
\beq
\sqrt{\varphi \over \ddot{\varphi}} \sim \nu^{-1}
= (2N)^{-1} (\Lambda_0 F''\lambda)^{-1/2}|_{\lambda=\varphi_M^2}
\sim {10^8 \over N} \cdot 300 {\rm yr},
\eeq
which is about $10^9$ years if $N \sim 10$.  
Here we assume $\varphi_M = \sqrt{\lambda} = O(1) = F''$ for systems of mild gravity
($\sim 10^{-8}$cm/s$^2$, e.g., clusters; for systems of stronger gravity,
the time scale is perhaps longer).
In the process of damping there will be a correction to MOND $\mu_M$ function
by the $q$ term, heuristically, $1- \varphi^2 = 1-\varphi_M^2$ if 
\beq
 x \rightarrow  \sqrt{{|\nabla \Phi|^2 \over \Lambda_0} + {q \over 2N^2\Lambda_0}},
\eeq
where $q \equiv \left[\partial_t^2 - c_\varphi^2 \nabla^2 + (1-c_\varphi^2)\eta
\partial_t\right]$ is an operator.  In tidally acting systems 
the value for $\varphi$ will oscillate between its pre-merging value and
its equilibrium value.

Models with a small $N$ would not give MOND.
E.g., if $N = 1-10$, $\pi \nu^{-1} \sim (100-10)$ Gyrs,
then the universe would be too young dynamically to have a precise MOND effect in galaxies
because $\varphi$ would not have enough time to respond to the formation of galaxies.  Rather $\varphi$ would lack behind, might remain close to its cosmological average:
\beq
\varphi \sim \bar{\varphi},
\eeq
which would mean a boost of the gravity of the baryon by a constant factor $(1-\bar{\varphi}^2)^{-1}$ everywhere.

\section{Generic Properties of Dark Fluid}

It is still uncertain whether the time-dependent correction and a possible diffusion term are enough to help MOND to explain the Bullet Clusters (Angus et al. 2007, Angus \& McGaugh 2008).
However, it seems robust that the {\it Dark Fluid} --
described by the field $Z^a=\AE^a \varphi(\lambda)$ -- is generally out of phase from the baryonic fluid.  There are two types of deviations from MOND in general:
\begin{itemize}
\item The Dark Fluid has a natural oscillation on time scales of $\nu^{-1}$,
which can be damped on a crossing timescale unless the external forcing is in resonance.
A very fast damping would mean an almost instantaneous relation between gravity and the scalar field $1-\varphi^2$, as the $\mu_M$ in classical Bekenstein-Milgrom (1984) modified gravity interpretation of MOND.  A slow damping would mean a history dependent relation, reminiscent of Milgrom's modified inertia interpretation of MOND: the dark fluid adds a dynamically-varying ineria around the baryons which it surrounds.
A possible test could be in galaxies with rotating bar(s), where there could
be a phase lag between the bar and the effective Dark Matter
(Debattista \& Sellwood 1998).  This has intriguing
consequences to the bar's pattern speed because of non-trivial corrections to the MOND pictures of dynamical friction (Ciotti \& Binney 2004, Nipoti et al. 2008, Tiret \& Combes 2008); the properties of the dark fluid is in bewteen that of real particle dark halo
and that naively expected from MOND.
\item The Dark Fluid has a pressure, controled by a propagation speed $c_\varphi$, 
where the speed of light is unity here, 
and the Dark Fluid can be made {\it Cold} by $c_\varphi^2 \sim 0$, or {\it Hot} by $c_\varphi^2 \sim 1$, or {\it Superluminal} by $c_\varphi^2 \ge 1$.  
The $\varphi$ would no longer be a function of the local gravity at ${\mathbf r}$
(as in MOND), rather it is a weighted average of a volume
of all points ${\mathbf r_1}$ by a Yukawa-type screening function
$\exp(- {\nu |{\mathbf r_1}-{\mathbf r}| \over c_\varphi })$, where
\beq
{\rm Screening~Length} = c_\varphi \nu^{-1} \sim c_\varphi {10^{8} \over N} \times ~{\rm 300 light~years}.
\eeq
Note this spatial correction to MOND can exist even in static systems;
even a small pressure term with
$c_\varphi^2 \ne 0$ might smooth out MOND effects on
small scale structures (wide binaries, star clusters, dwarf galaxies),
where the wavenumber $|{\mathbf k}|^2$ is much bigger than in galaxy clusters.
The screening length can be set at $\sim 100$ pc for either a $N\sim 10^8$, 
$c_\varphi \sim 3\times 10^5$ km/s Hot Dark Fluid {\it or} a $N\sim 10^4$ 
and $c_\varphi \sim 30$ km/s Cold Dark Fluid.  
This scale 100pc is a scale dividing dense star clusters and fluffy dwarf galaxies.
Observationally dark matter effects are only seen in the universe on scales larger than 100pc.  It has been challenging for MOND to explain this observed scale (Zhao 2005, Sanchez-Salcedo \& Hernandez 2007, Baumgardt et al. 2005).
\end{itemize}

In conclusion, we find a framework of Dark Fluid theories where
MOND corresponds a special choice of potentials or mass for the
vector field.  The Dark Fluid can run Cold or Hot depending on the sound speed 
$c_\varphi$ (which could even be a running function of the vector field).
These theories degenerate into scalar field theories
for Dark Energy effects in the Hubble expansion. It is possible to
create an exact $w=-1$ Dark Energy effect (and a Dark Matter effect
for $b^{-1} = {\varphi^2 \over \varphi^2-1} >1$).  The scale $a_0
=\sqrt{\Lambda_0}$ in MOND in equilibrium spiral galaxies derives its physics from the
amplitude of the dark energy $\Lambda_0$.
MOND or Dark Matter effects are hence indications of a
non-uniform Dark Energy fluid described generally by a vector
field $Z^a$.  For non-equilibrium systems like the Bullet Clusters or
galaxies with satellites, the properties of the Dark Fluid
do not follow exactly the usual expectations of MOND or Cold/Hot Dark Matter,
but (not so surprisingly) in between.

\acknowledgements
HSZ acknowledges Xufen Wu for assistance in making the figures.

\begin{figure*}
\plotone{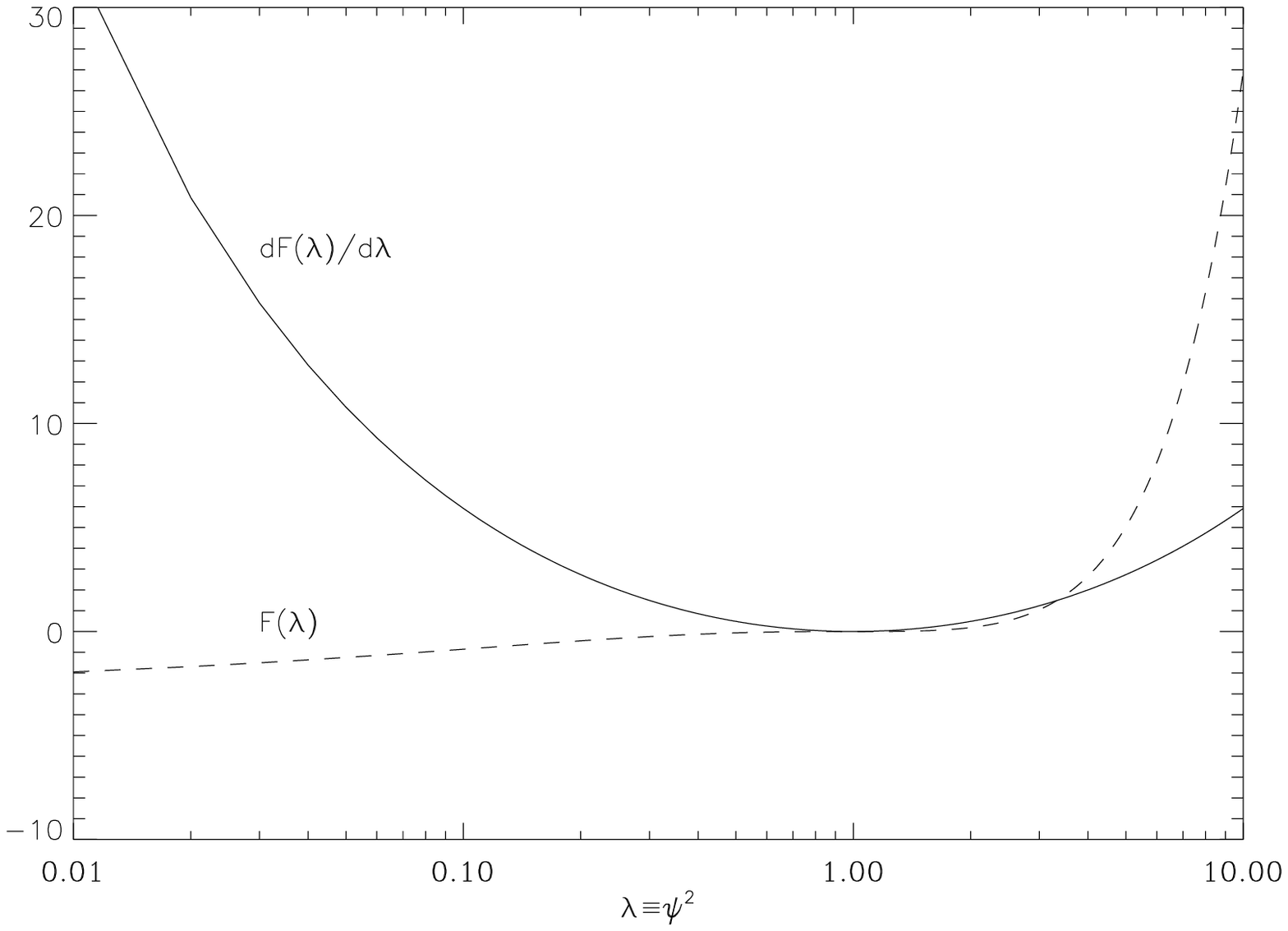}
\plotone{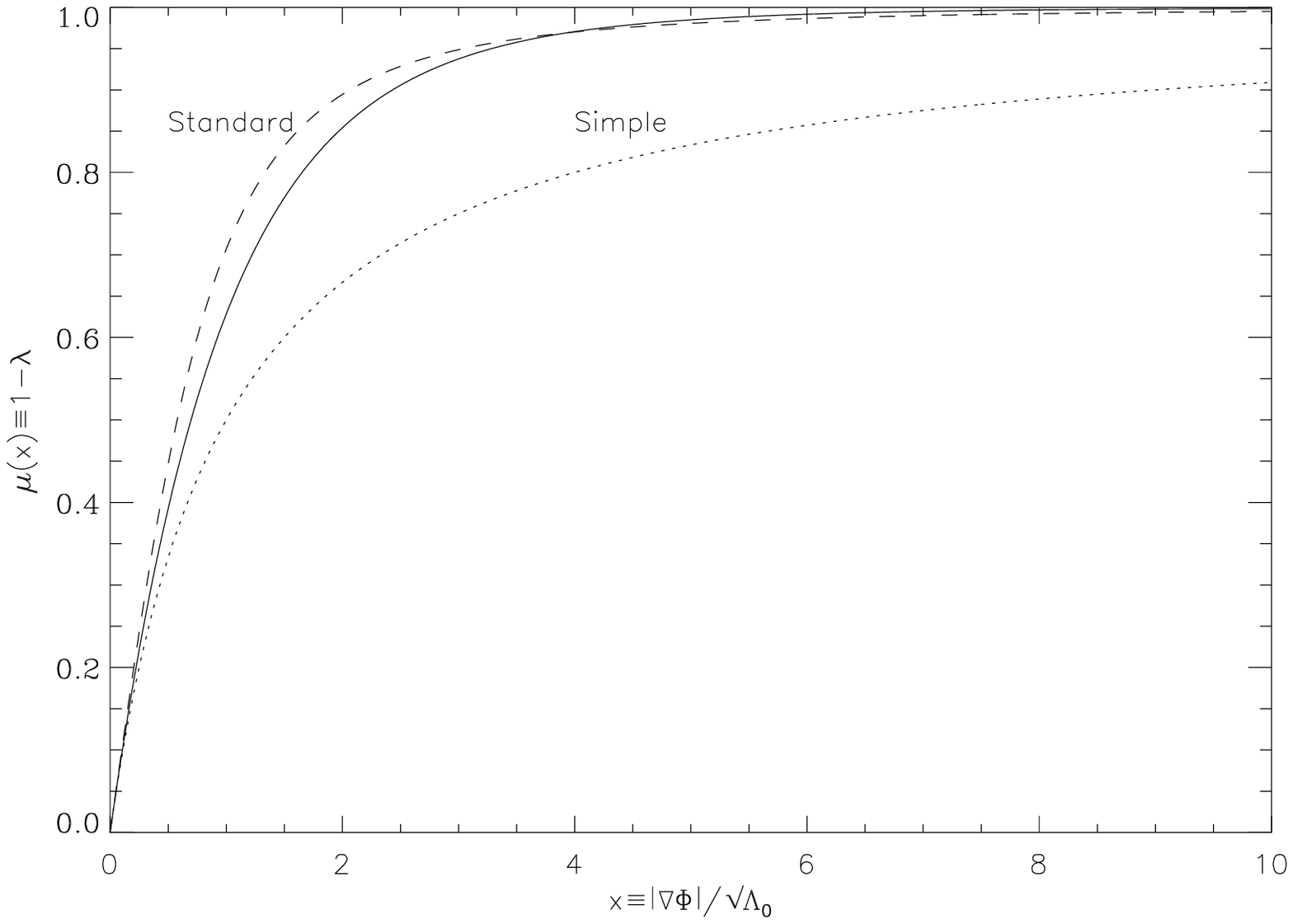}
\caption{
Panel (a) shows $F(\lambda)={8(\varphi-1)^3/3}$ (dashed) and 
${d\over d\lambda}F=4(\varphi-1)^2 \varphi^{-1}$ (solid) 
as functions of $\lambda \equiv \varphi^2$.
Panel (b) shows our function $\mu \equiv 1-\varphi_M^2 
=1-\left(\sqrt{1+ {x^2 \over 16}}+{x\over 4} \right)^{-4}$ as function of 
$x \equiv |\nabla \Phi|/\sqrt{\Lambda_0}$ (solid).  Overplotted is the MOND 
$\mu={x/\sqrt{1+x^2}}$ (dashed, labeled "standard") 
and the MOND $\mu={x/(1+x)}$ (dashed, labeled "simple"), adopting 
$\sqrt{\Lambda_0} \rightarrow a_0$. 
}
\end{figure*}

\appendix 
\section{Appendix: the vector field equation and the damping rate $\eta$}

Consider the vector field perturbation
$\ae_j=Y_{,j} \ll 1$ in the compressional spin-0 mode with a potential $Y$.   
Define $C_2= {c_2 \over c_4}={b-1 \over 3\lambda}$, we note 
the vector field EOM in the raw form is 
\beq
\partial_t
\left[ 2 \lambda (\dot{Y}_{,i} + A_i ) \right] -
\partial_i \left[2C_2\lambda\left(Y_{,jj} + \theta \right) \right] = - {n \over N^2} \dot{\varphi}\varphi\left[
{\varphi_{,i} \over \varphi} - { \dot{\varphi} \over \varphi}
Y_{,i} \right]
\eeq 
for the index $i=x,y,z$, where $n=1-c_\varphi^2$, and 
$A_i \equiv u^a \nabla_a u_i$ and 
$\theta \equiv \nabla_a u^a$, and $u^a$ is the unit four-velocity vector locally.
Apply the approximation $\lambda$ is a very small constant (strong gravity), 
so that $A_i \sim -\Phi_{,i}$ 
and $\partial_i \theta \sim -3 \dot{\Psi}_{,i}$, and neglect the term
${\dot{\varphi} \over \varphi} Y_{,i}$ because $|Y_{,i}|=|\ae_i|
\ll 1$ for perturbations, and $|\dot{\varphi}| \ll
|\dot{\varphi}_{,i}|$ inside the causual horizon. Replace
$\varphi=\sqrt{\lambda}$ and apply $\partial_i$ to both sides,
we get 
\beq
\partial_i \partial_t \left[ 2\lambda (\dot{Y}_{,i} -
\Phi_{,i}) \right] - \partial_i
\partial_i \left[ 2C_2 \lambda (Y_{,jj} - 3
\dot{\Psi} ) \right] = -\partial_i \left[ {n \dot{\lambda} \over
4N^2\lambda} \lambda_{,i} \right] \sim O\left({ n \over 4N^2}
\right) |{\mathbf k}|^2 \omega \lambda .
\eeq
Furthermore, we neglect
the spatial and temporal variations of $\lambda$ and $C_2$,
factor out $2\lambda$, and define
\beq
 \eta \equiv \sum_{j=x,y,z} \ae_{j}^{,j} = -\nabla^2 Y = |{\mathbf k}|^2 Y,
\eeq 
then the approximate equation for $\eta$ is obtained:
\beq
\left[\partial_t\partial_t - C_2 \nabla^2 \right] \eta \sim
-\nabla^2 \dot{\Phi} + 3C_2 \nabla^2 \dot{\Psi} + O({n
\omega \over 4N^2}) |{\mathbf k}|^2 \sim O(\omega^3) \left[1+ {n
\over 4} O\left([N\sigma]^{-2}\right) \right], \eeq where $n \equiv 1-c_\varphi^2$
and $C_2$ plays the role of sound speed squared.  
Replace $\Psi=-\Phi$, and replace the partial derivative $\partial_i$ with the wave
vector ${\mathbf k}$, and $\partial_t$ with the orbital frequency
$\omega$, we get 
\beq \eta \sim {(1+3C_2)|{\mathbf k}|^2
\dot{\Phi} \over -\omega^2 + C_2 |{\mathbf k}|^2 } \sim
{1+3C_2 \over 1 - C_2\sigma^{-2} } \times \omega
\sim O([{\rm orbit~crossing~time}]^{-1}). 
\eeq 
This estimation is
good for fairly hot system $\sigma \gg 1/N \sim $
30km/s if $N \sim 10^4$, and $\sigma \gg \sqrt{|C_2|}$.
This is an over-estimation if the vector field has a
large (relativistic) sound speed due to a not-so-small $\sqrt{|C_2|}$;
this is an under-estimation if the vector field has a small sound speed
$\sqrt{|C_2|}$ in resonance with the stellar velocity $\sigma$ or if
the system is very cold $\sigma \le 1/N$.  There will also be
corrections of order $H/\omega$ in an expanding universe.  

{}

\end{document}